\documentstyle[prl,aps,epsf,multicol]{revtex}

\input{psfig}

\begin{document}

\date {DART-HEP-96/01 $~~~~$ UFIFT-HEP-96-2 $~~~~$ January 1996}

\vspace{0.5in}
\title{Oscillons in a Hot Heat Bath}

\vspace{1.cm}

\author{
Marcelo Gleiser$^1$\footnote{NSF Presidential Faculty Fellow}\footnote{
email: gleiser@peterpan.dartmouth.edu}
and Richard M. Haas$^2$\footnote{email: rhaas@phys.ufl.edu} }

\vspace{1.0cm}

\address{ $^{1)}$ Department of Physics and Astronomy,
Dartmouth College,
Hanover, NH 03755}

\address{ $^{2)}$ Department of Physics,
University of Florida,
Gainesville, FL 32611}

\maketitle

\vspace{1.cm}

\begin{abstract}
In models of real scalar fields with degenerate double-well
potentials, spherically symmetric, large amplitude
fluctuations away from the vacuum are unstable.
Neglecting interactions with an external environment, the evolution
of such configurations may entail the development
of an oscillon;
a localized, non-singular, time-dependent configuration which is
{\it extremely} long-lived.
In the present study we investigate numerically 
how the coupling to a heat bath influences the evolution of collapsing bubbles.
We show that the existence and lifetime of the
oscillon stage is extremely sensitive to how strongly the field is coupled
to the heat bath. By modeling the coupling through a Markovian Langevin
equation with viscosity coefficient $\gamma$, we find that for
$\gamma \gtrsim 5 \times 10^{-4}m$, where $m$ is the
typical mass scale in the model, 
oscillons are not observed.

\vspace{0.5cm}


\end{abstract}
\newpage

\begin{multicols}{2}
\narrowtext

\section{Introduction}

It is well-known that 
nonlinear field theories allow for the existence of static, regular,
and localized configurations \cite{SOLITON}. In (1+1)-dimensions,
one finds the kink solutions to the Klein-Gordon equation with sine-Gordon
or $\phi^4$ potentials. Unfortunately, for a higher number of spatial
dimensions, the only static and stable solutions involve either two or
more fields with some topological conservation law, as in the case of
the 't Hooft-Polyakov monopole or the Nielsen-Olesen vortices
\cite{RAJARAMAN}, or a conserved global charge, as in the case of 
nontopological solitons \cite{NTSs}.
The interest in such static solutions stems from a myriad of possible
applications, from the modeling of particles 
and cosmological topological defects, to the propagation of 
information in optical fibers and defects in liquid crystals and superfluids.

Given the justified interest in static solutions, unstable,
time-dependent solutions
have received considerably less attention in the literature. 
However, as
recent work has emphasized, the presence of
nonlinearities can lead to the existence of extremely long-lived ({\it i.e.},
nearly non-dissipative) solutions
of the Klein-Gordon equation \cite{MG,CGM}. 
These spherically symmetric solutions, known as oscillons,
are characterized by a rapid oscillation of the field at
small radial values (the core) of the configuration,
very much like the (1+1)-dimensional breathers which form
from kink-antikink collisions \cite{BREATHERS}. In fact, oscillons can be 
thought of as a possible stage during the collapse of an unstable bubble;
as the bubble collapses radiating its initial energy to infinity, it may
(or may not) settle into an oscillon before completely
disappearing. 
[The detailed conditions for a bubble to settle into an oscillon
configuration are given in Ref. \cite{CGM}.]

Apart from adding to our knowledge of time-dependent coherent nonlinear
phenomena in field theories, the fact that collapsing bubbles may settle
into long-lived configurations may have important consequences for
our understanding of the dynamics of phase transitions.
As an example, consider a system in a metastable state which is
being cooled down at some rate per unit volume, $\Gamma_{\rm cool}$.
For a strong enough first order phase transition, where homogeneous
nucleation theory is applicable, the system will supercool
in this metastable state before it decays by nucleating a critical radius
bubble. The rate per unit volume for nucleating a critical bubble is dominated
by the Boltzmann factor, 
$\Gamma_{\rm cb} \sim {\rm exp}[-F_{\rm cb}/T]$,
where $F_{\rm cb}$ is the cost in free energy for nucleating
the critical bubble, and $T$ is the temperature.
Critical bubble nucleation is suppressed
for $\Gamma_{\rm cb}/\Gamma_{\rm cool} < 1$.
Unless the barrier disappears below a certain temperature,
the
free energy cost for nucleating a critical bubble typically decreases
as the system is cooled down, 
reaching a minimum value before it starts increasing again \cite{GUTH-WEINBERG}.
Apart from critical bubbles, smaller free energy,
unstable (subcritical) bubbles are also nucleated. Since subcritical
bubbles may fall
into an oscillon stage during their collapse, they may exist
for a much longer time than one would naively estimate. If their lifetime,
$\tau_{\rm osc}$, is longer than the cooling time-scale 
$[ \tau_{\rm osc} > (\Gamma_{\rm cool}V)^{-1} ]$,
it is possible that as the temperature decreases,
they will overcome the free-energy barrier
and initiate the decay of the metastable state. In other words, oscillons 
could become critical bubbles. 

In practice, this possibility has not yet 
been analysed in much detail. Estimates
in the context of the electroweak phase transition show that oscillons are
sufficiently suppressed to be of negligible impact \cite{Riotto}. However,
based on the qualitative arguments above, one expects oscillons to be
of greater relevance in earlier (in a cosmological context), 
and thus faster, phase transitions. In fact,
in the work of Copeland, Gleiser, and M\"uller, it was remarked that
for a GUT-scale transition
oscillons are efficiently produced by thermal processes if 
$F_{\rm osc}/T < 10$. This condition may be satisfied by, e.g.,
Coleman-Weinberg potentials. Also, apart from their potential
relevance to the dynamics of both cosmological and laboratory
phase transitions, 
it remains to be seen if oscillons
exist in the non-relativistic limit of the Klein-Gordon equation, as 
time-dependent solutions of the nonlinear Schr\"odinger equation.

The above discussion neglected the effects of an
external environment on the evolution of shrinking bubbles. 
However, in most realistic
situations, a thermal background will influence the 
dynamics (and coherence) of any field configuration. 
In this work, we will examine the evolution of coherent field
configurations coupled to 
heat baths in more detail. The evolution of the scalar field in a 
thermal bath will be modelled by a Markovian Langevin equation. As we
will see, the presence of a thermal bath adds additional constraints 
on the possible existence and lifetimes of oscillons; 
even if the collapsing 
bubble does settle into an oscillon, the bath can affect its
lifetime quite appreciably. However, for sufficiently small couplings between
the field and the thermal bath, oscillons are still present.

The rest of this work is organized as follows.
In the next section, we briefly review the properties of oscillons in the
absence of a thermal bath.
In Section III, we discuss how we study
bubble evolution in the presence of a heat bath.
In Section IV, we present 
our numerical results. Lifetimes of the oscillons for several sets of
parameters are obtained using
a method consistent with an ensemble-averaging procedure,
together with empirically determined equations 
of best fit for the data. In Section V we summarize our results, pointing
to possible directions of future work.

\section{Cold Oscillons}

In this section we briefly review some of the
results which established the existence
of oscillons as a possible stage during bubble collapse for nonlinear
field theories. This discussion will help us set up the notation and
concepts which will be useful later when we include thermal effects.
More details can be found in Refs. \cite{MG,CGM}. 

The action for a real scalar field in (3+1)-dimensions is
\begin{equation}
S[\phi] = \int d^4x \left[ {1\over 2} (\partial_{\mu} \phi)
(\partial^{\mu} \phi) - V(\phi) \right ] ~~,
\end{equation}
\noindent 
where we restrict our investigation to a symmetric double-well potential
of the form
\begin{equation}\label{potsdwp}
V(\phi) = {\lambda\over 4} \left( \phi^2 -
 {m^2\over\lambda} \right)^2.
\end{equation}

A solution $\phi({\bf x},t)$ to the equation of motion,
\begin{equation}\label{fullEoM}
\partial^2\phi/\partial t^2 -
\nabla^2 \phi = - {\partial V(\phi)\over \partial \phi} ~~,
\end{equation}
\noindent
has energy
\begin{equation}
E[\phi] = \int d^3x \left[ {1\over2} (\partial\phi/\partial t)^2 +
{1\over 2} (\nabla \phi) ^2 + V(\phi) \right] ~~.
\end{equation}

Since we are only interested in spherically-symmetric configurations,
it proves convenient to introduce dimensionless variables,
$\rho=rm,~\tau=tm,~{\rm and}~\Phi={{\sqrt{\lambda}}\over m}\phi$. The nonlinear
Klein-Gordon equation is,
\begin{equation}\label{eqmotsp}
{{\partial^2\Phi}\over {\partial\tau^2}}-{{\partial^2\Phi}
\over {\partial \rho^2}}
-{2\over {\rho}}{{\partial \Phi}\over {\partial \rho}}  = \Phi - \Phi^3.
\label{equ_of_mot:eq}
\end{equation}

Oscillons are found by solving the above equation with a bubble-like
initial configuration, which so far has been taken to be either a
``Gaussian'' (equ. 6) or a ``tanh'' (equ. 7) bubble, 
\begin{eqnarray}\label{initg}
&\Phi(\rho,0)& =
(\Phi_c - \Phi_0)\, e^{-\rho^2/R_0^2} + \Phi_0 \\
\label{initt}
&\Phi(\rho,0)& ={1\over 2}\left[ (\Phi_0 - \Phi_c)\,
\tanh (\rho - R_0) +
\Phi_0 + \Phi_c \right]~,\\
& & \hspace{2.1in} R_0\gg 1~. \nonumber
\end{eqnarray}
$R_0$ is the initial radius of the configuration, $\Phi_0$ is the asymptotic
vacuum the configuration decays into (we will take $\Phi_0=-1$), and
$\Phi_c$ is the value of the field at the configuration's core. [We will
take it to be $\Phi_c=1$, {\it i.e.}, the bubble interpolates between the
two vacua.] We also impose the boundary conditions,
\begin{equation}\label{bc}
\Phi(\rho\rightarrow \infty,\tau) = \Phi_0,\; \Phi'(0,\tau) = 0,\;
\dot\Phi(\rho,0) = 0\, .
\end{equation}

\psfig{file=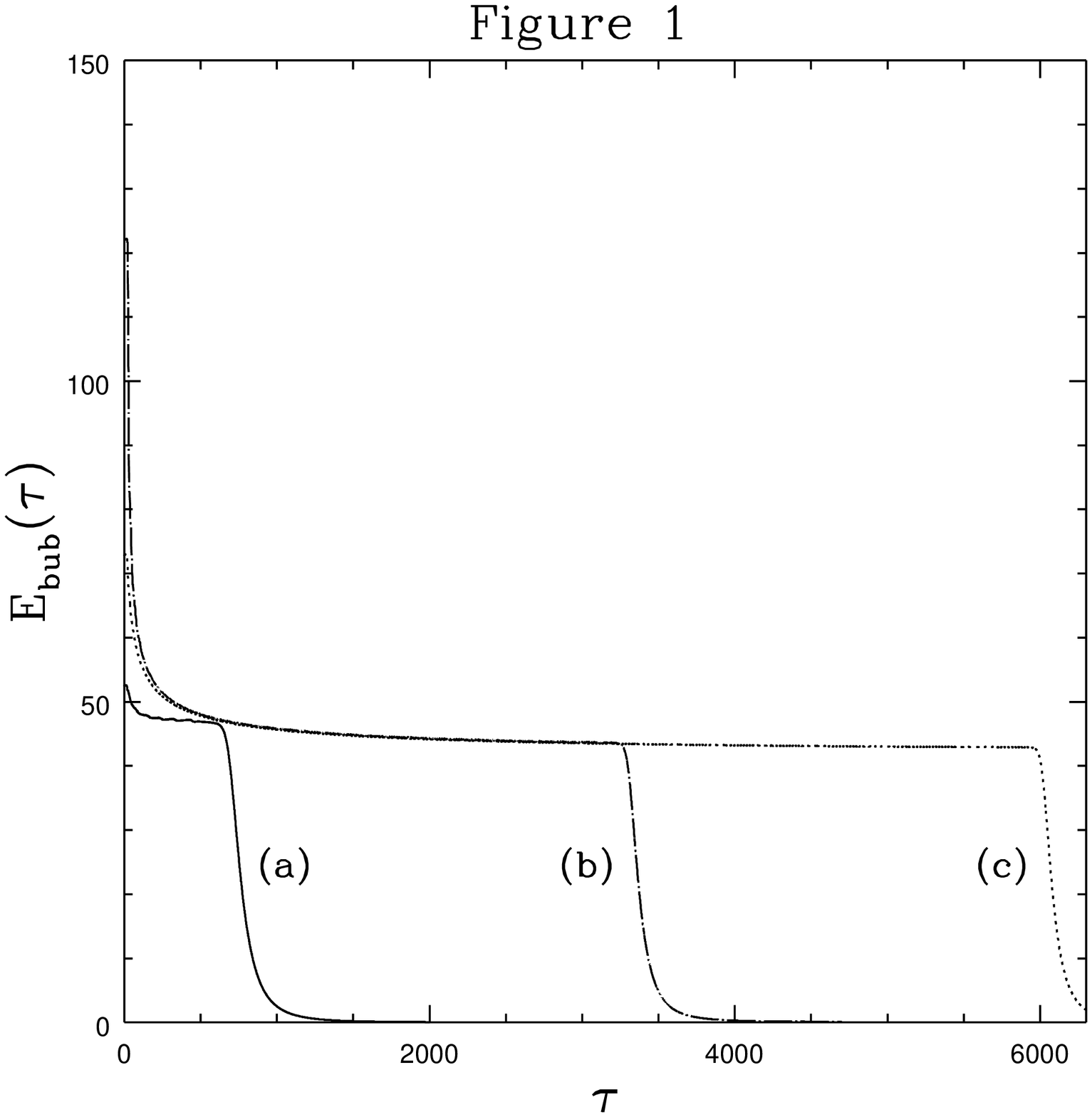,width=234pt}
\begin{figure}
	\caption{
		Bubble energy vs. time for
		$\gamma = 0$ and $T = 0$ in the case of
		(a) $R=2.3$, (b) $R=3.4$, and (c) $R=2.7$.
	}
\end{figure}

Equation \ref{equ_of_mot:eq} is then numerically 
solved by a finite difference routine second
order accurate in time and fourth order accurate in space. 
[We used the same discrete steps as in Ref. \cite{CGM}, that is, a spatial
step $h=0.1$ and a time step $\delta\tau=0.05$.]
In order to examine
the rate at which the initial configuration decays into the vacuum, we
measure the
total energy within a spherical shell surrounding it.
As discussed in Refs. 
\cite{MG,CGM}, the oscillon configuration is identified by the presence of
an almost flat plateau in the plot of energy vs. time, signalling a regime
during which almost no energy is radiated away. Typically, lifetimes
range from $10^3m^{-1}$ to $10^4m^{-1}$. In Figure 1, we show the 
energy of a few oscillons as a function
of time for several Gaussian bubbles of differing initial radii.

\section{Hot Oscillons}

We would like to extend the previous analysis for fields interacting with
thermal baths. The simplest way to do this is to couple the scalar field to
a thermal bath via a generalized Langevin equation, which  assumes the bath
to be Markovian (white noise) and that the field couples to the noise
additively. The bath is characterized by a viscosity coefficient $\gamma$
and by a random noise $\xi({\bf x},t)$, which are related by the 
fluctuation-dissipation theorem,
\begin{equation}
\langle \xi({\bf x},t)\xi( {\bf x'},t')\rangle =
  2 \gamma T \delta(t-t')\delta^3({\bf x} - {\bf x'})~~.
\end{equation}

$\gamma$ represents the coupling strength between field and thermal bath.
For simple models, it can be expressed perturbatively in terms of the
coupling constants in the system. [See, e.g., Ref. \cite{GR}.] However, we will
treat it here as a free parameter. Note that
$\gamma^{-1}$ defines the thermalization time-scale.
More complicated nonlocal forms of the Langevin equation could be
used, although we think it wise to defer such approaches to later studies.
[More details can be found in Refs. \cite{GR,BG} and references therein.]

In terms of the dimensionless coordinates defined before, the 
Langevin equation is given by,

\begin{equation}\label{langevin}
{{\partial^2\Phi}\over {\partial\tau^2}} + {\tilde \gamma}{{\partial\Phi}\over
{\partial \tau}}
-{{\partial^2\Phi}
\over {\partial \rho^2}}
-{2\over {\rho}}{{\partial \Phi}\over {\partial \rho}}  = \Phi - \Phi^3
+ {\tilde \xi},
\end{equation}
where ${\tilde \gamma}=\gamma/m$ and ${\tilde \xi}=\sqrt{\lambda}\xi/m^3$
are the dimensionless viscosity and noise respectively.

Here, one last simplification 
comes in. In general, one should be solving the fully (3+1)-dimensional
Langevin equation, as the bath carries no symmetries with it. However, we will
assume that the dominant interactions of the bubble with the thermal bath are
in the radial direction. We must rewrite the fluctuation-dissipation
relation consistently with  spherical symmetry, effectively keeping the problem
one dimensional. The full (3+1)-dimensional problem proves to be 
very CPU intensive and of limited interest. 
[Note that in the presence of a
thermal bath, results are obtained after an ensemble averaging
procedure. In other words, each measurement can involve hundreds of
runs. As it is, the results obtained here already required a network
of workstations running for a total of 6000 days of CPU time.]
This approximation is not only economical, but also makes sense
physically.
For bubble-like configurations, it is reasonable
to expect that the effect of the bath will be dominant in the radial
direction. In any case, the results here should be considered as an upper
bound on the lifetimes; including perturbations in all three dimensions
will not help increase the oscillon's lifetime.

The spherically-symmetric fluctuation-dissipation relation is,
\begin{equation}
\langle {\tilde \xi}(\rho,\tau){\tilde \xi}(\rho^{\prime},\tau')\rangle =
  {1\over {2\pi\rho^2}}
{\tilde \gamma} \theta \delta(\tau-\tau')\delta(\rho - \rho^{\prime})~~,
\end{equation}
where $\theta =\lambda~T/m$ is the dimensionless temperature.

Before proceeding, we must make sure that our expression for the 
fluctuation-dissipation relation is physical. A simple way to do this is to
consider the evolution of a free field in a parabolic potential and measure the
energy per degree of freedom $E/N$, where $N$ is the total number of
spatial lattice points. Since the problem is one dimensional, 
the equipartition theorem states that in equilibrium, $E/N = T/2$. We have
confirmed that this relation is satisfied for a wide range of parameters,
as shown in Figure 2.

\psfig{file=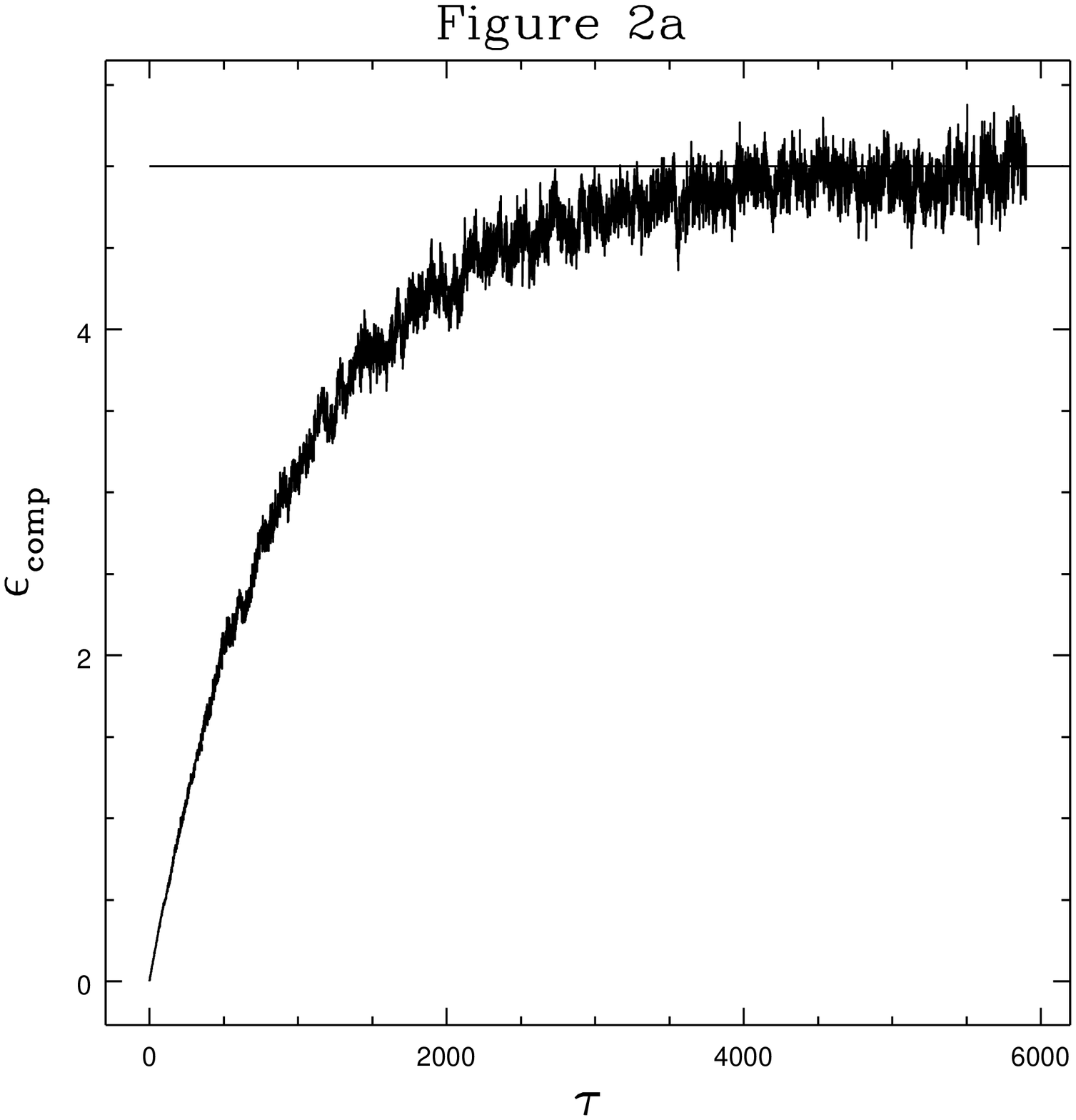,width=204pt}
\psfig{file=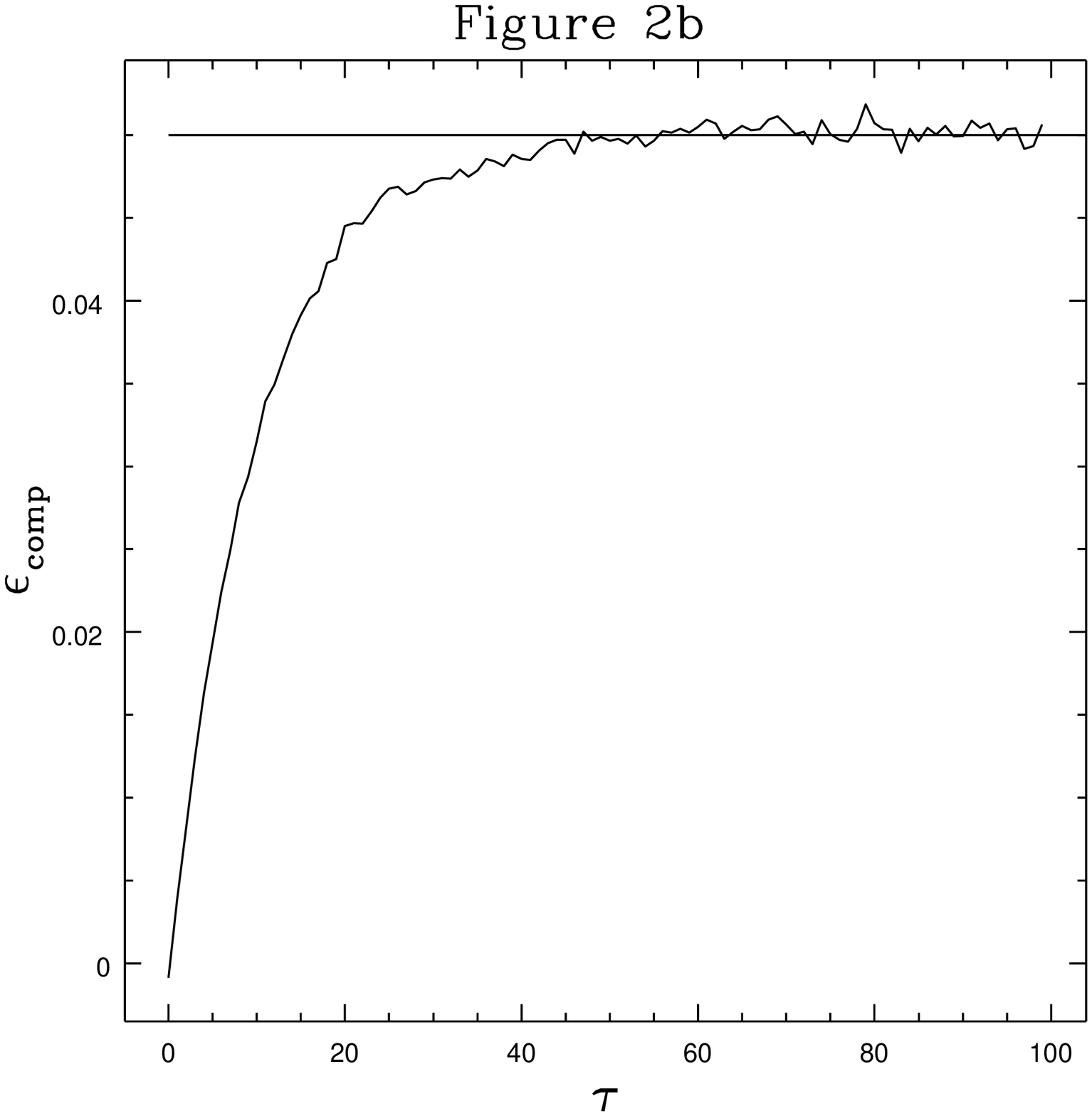,width=204pt}
\psfig{file=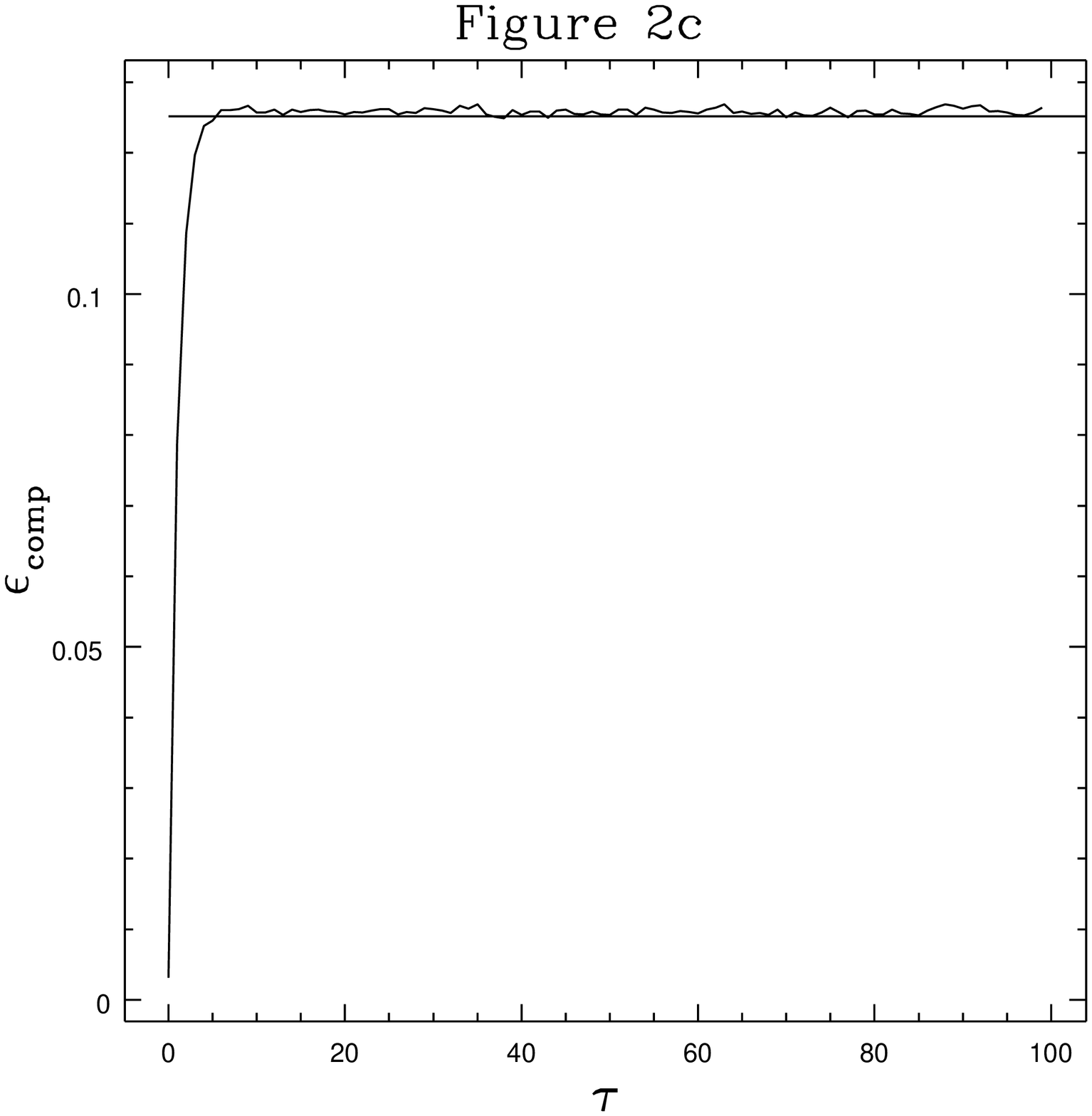,width=204pt}
\begin{figure}
	\caption{
		Computed energy per degree of freedom, 
		$\epsilon_{comp}$, vs. time for
		(a) $\gamma = 10^{-3}$ and $T=10.0$,
		(b) $\gamma = 0.1$ and $T=0.1$,
		(c) $\gamma = 1.0$ and $T=0.25$.
		The horizontal lines in each figure represent the
		expected values from the equipartition theorem. 
	}	
\end{figure}

\section{Numerical Results}

Now that we have an evolution equation in the presence of a thermal
bath consistent with the assumption of spherical symmetry, we can start
investigating how the bath influences the formation and longevity of
oscillons.

As in the ``cold oscillon'' case, we measure the energy within a shell
surrounding the initial configuration as a function of time. The radius
of the shell, $R_{\rm shell}$, should be taken to be sufficiently larger than
the radius of the initial configuration, $R_0$. Typically, we chose
$R_{\rm shell}=15$.
The presence of
temperature, however, immediately introduces two difficulties.
First, the bath contributes to the energy within the shell, obscuring the
measurement of the bubble's energy. Second, a physical result must be
obtained by an ensemble averaging procedure, where different conditions are
set by different seeds for the random number generator responsible for the 
noise term in the equation. However, as soon as we change the seed, the oscillon
lifetime changes, as we show in Figure 3 for a selection of 30 runs out
of 150.
We must find a consistent procedure for measuring the
ensemble-averaged oscillon lifetime. Let us deal with each of these problems
in turn.

\psfig{file=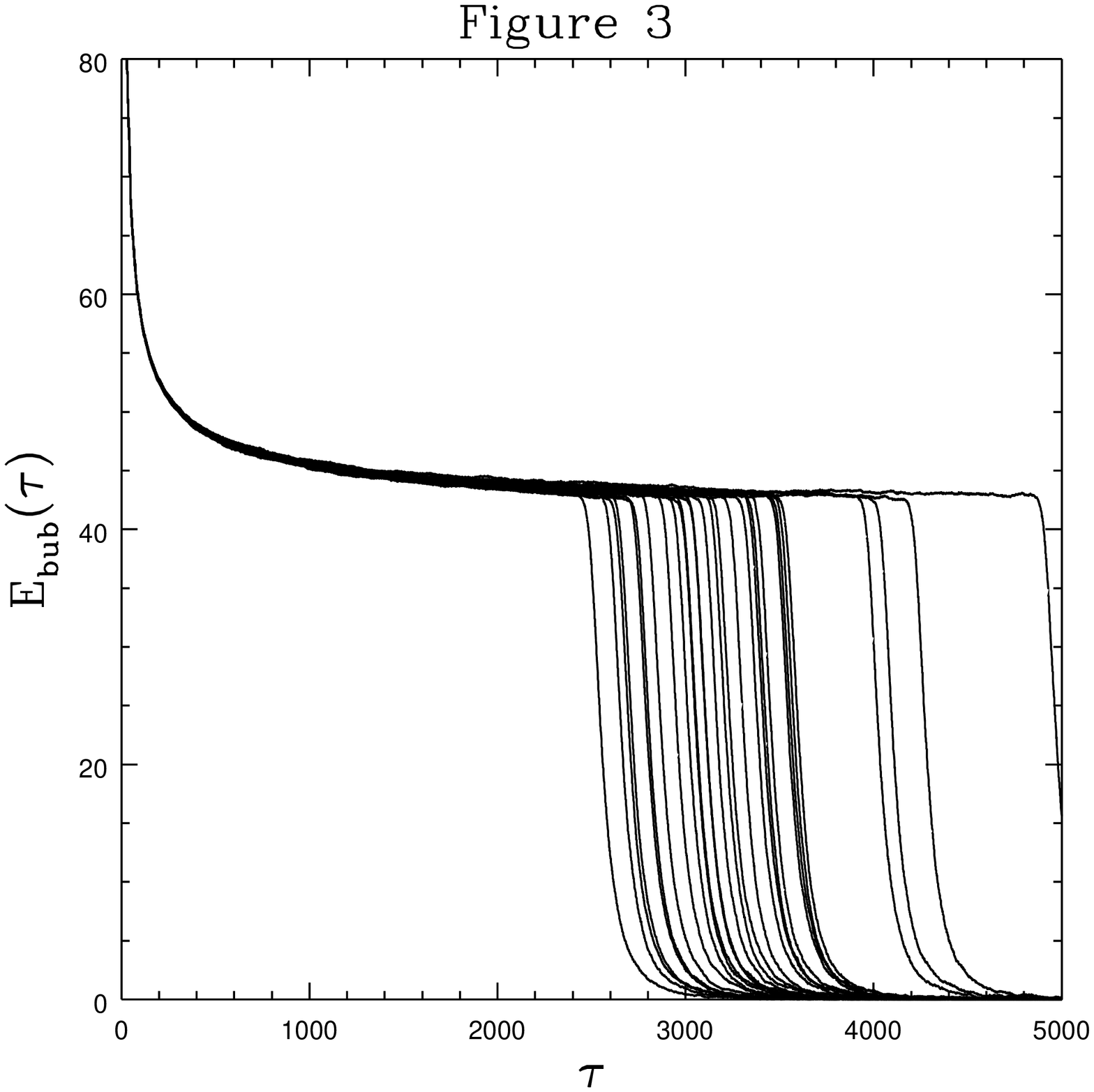,width=234pt}
\begin{figure}
	\caption{
		Bubble energy vs. time for 
		$R=3.2$, $\gamma=10^{-5}$, and $T=0.1$ with
		different seed values for each curve.
		The figure represents only $30$ runs from a total of $105$.
	}
\end{figure}

The first problem can be taken care of quite easily; we perform several
runs without a bubble [{\it i.e.}, with a constant initial configuration,
$\Phi(\rho,0)=-1$], 
and measure the ensemble-averaged
energy of the bath within the shell, $\langle E_{\rm bath} \rangle$. 
The angular brackets denote ensemble-averaged quantities. We
then define the normalized energy of the bubble by
\begin{equation}
E_{\rm bub}=E_{\rm total} - \langle E_{\rm bath} \rangle~~,
\end{equation}
where, $E_{\rm total}$ is the total energy within the shell. 
Figure 4 shows a typical case for 
$E_{\rm bub}$, $E_{\rm total}$, and $\langle E_{\rm bath} \rangle$.
As a test of this procedure, we confirmed that 
$\langle E_{\rm bub}\rangle \rightarrow 0$
after bubbles have decayed. Another test is to measure the plateau energy
for small $\gamma$ and $T$. The result should be (and is) consistent with the
cold oscillon case, $E_{\rm plateau}\sim 43m/\lambda$.

\psfig{file=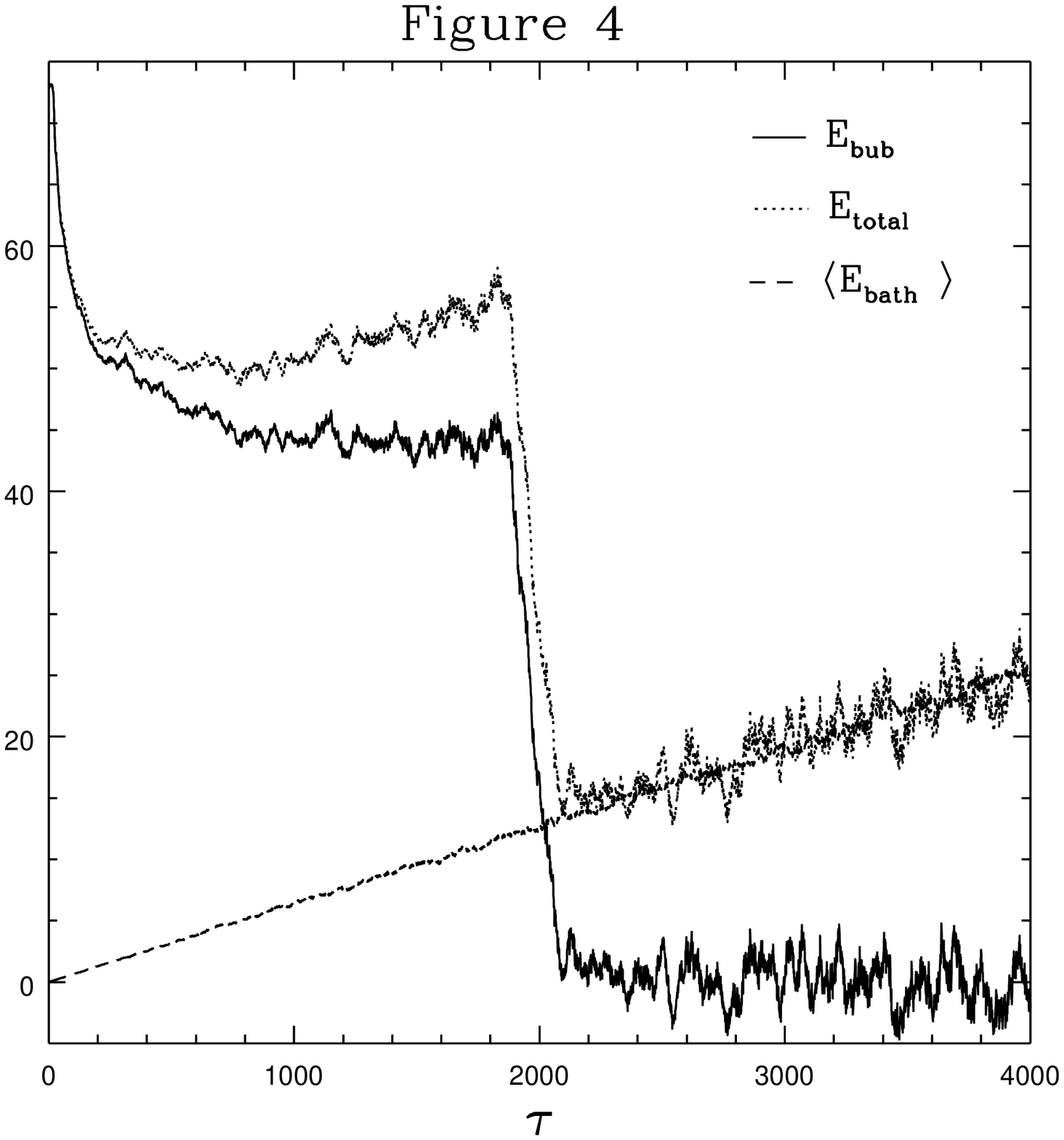,width=234pt}
\begin{figure}
	\caption{
		Bubble energy, total energy, and bath energy as 
		functions of time for $R_{0} = 2.7$, $\gamma = 10^{-5}$,
		and $T=5.0$.
	}
\end{figure}

In order to handle the second problem, we first define how we measure the
oscillon's lifetime. As opposed to the case of cold oscillons, where we
could just estimate the lifetime by checking when the oscillon finally 
decays (see Figure 3),
here we must automate the procedure, as we will have to handle thousands of
runs. First, we must establish if a given initial bubble settles into an
oscillon stage. As we remarked earlier, the most important signature of the
oscillon is the near constancy of its total energy. In Refs. \cite{MG,CGM},
it was measured that during the oscillon stage the plateau energy varies
by at most $20\%$. After an exhaustive search,
we found that we can identify an oscillon stage by setting
$|dE_{\rm plateau}/dt|\leq 0.02m^2/\lambda$. 
Steeper changes of the plateau energy were incompatible with the nearly
non-dissipative nature of oscillons.
Now that we know how to hunt for oscillons, we must decide how we will measure
their lifetimes. Here, the fact that these configurations decay rather quickly
allows us to simply set the oscillon lifetime to the time when
its energy drops to $0.1E_{\rm plateau}$.

To summarize so far,  we first choose values for
the three variables in the model: the initial radius $R_0$, the
viscosity $\gamma$, and the temperature $T$. Each numerical
``experiment'' consists of
evolving the Langevin equation with a choice of variables and a
given value for the random number seed.
The code then searches for an oscillon by
identifying a fairly flat plateau, and then measures its lifetime by
recording the time when its energy decays to $0.1E_{\rm plateau}$. 
This procedure
is repeated a large number of times with different seeds for the random
number generator. This way, we obtain a distribution of lifetimes for a
given set of variables, as shown in Figure 3. In order to determine the
ensemble-averaged lifetime, we produce a histogram where we bin different
decay times vs. the number of occurrences. In Figure 5, we show a few examples
of histograms
for varying radii and fixed viscosity and temperature. In Figure 6, 
we show a few examples of varying viscosity, with fixed radius and temperature.

\psfig{file=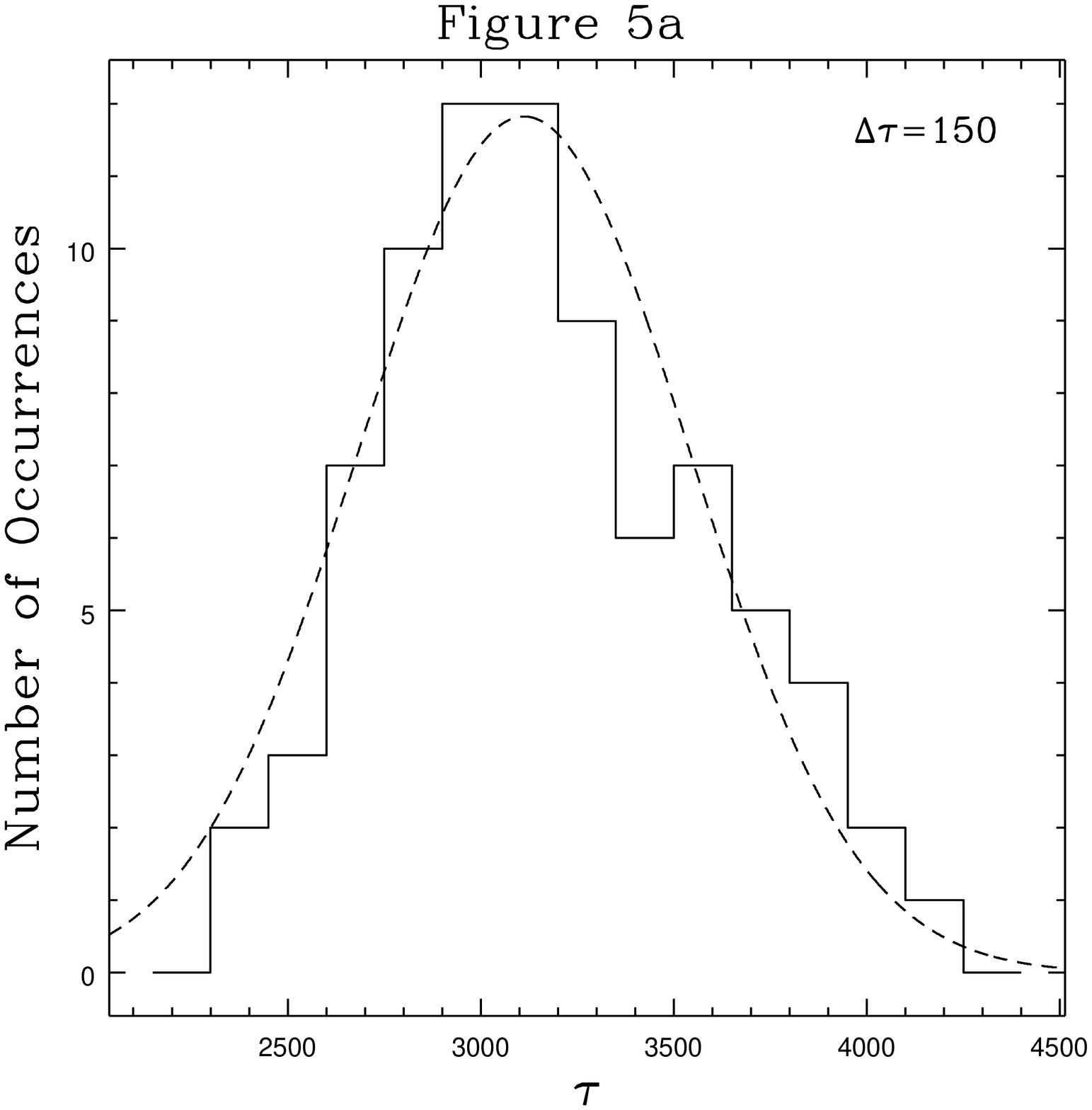,width=228pt}
\psfig{file=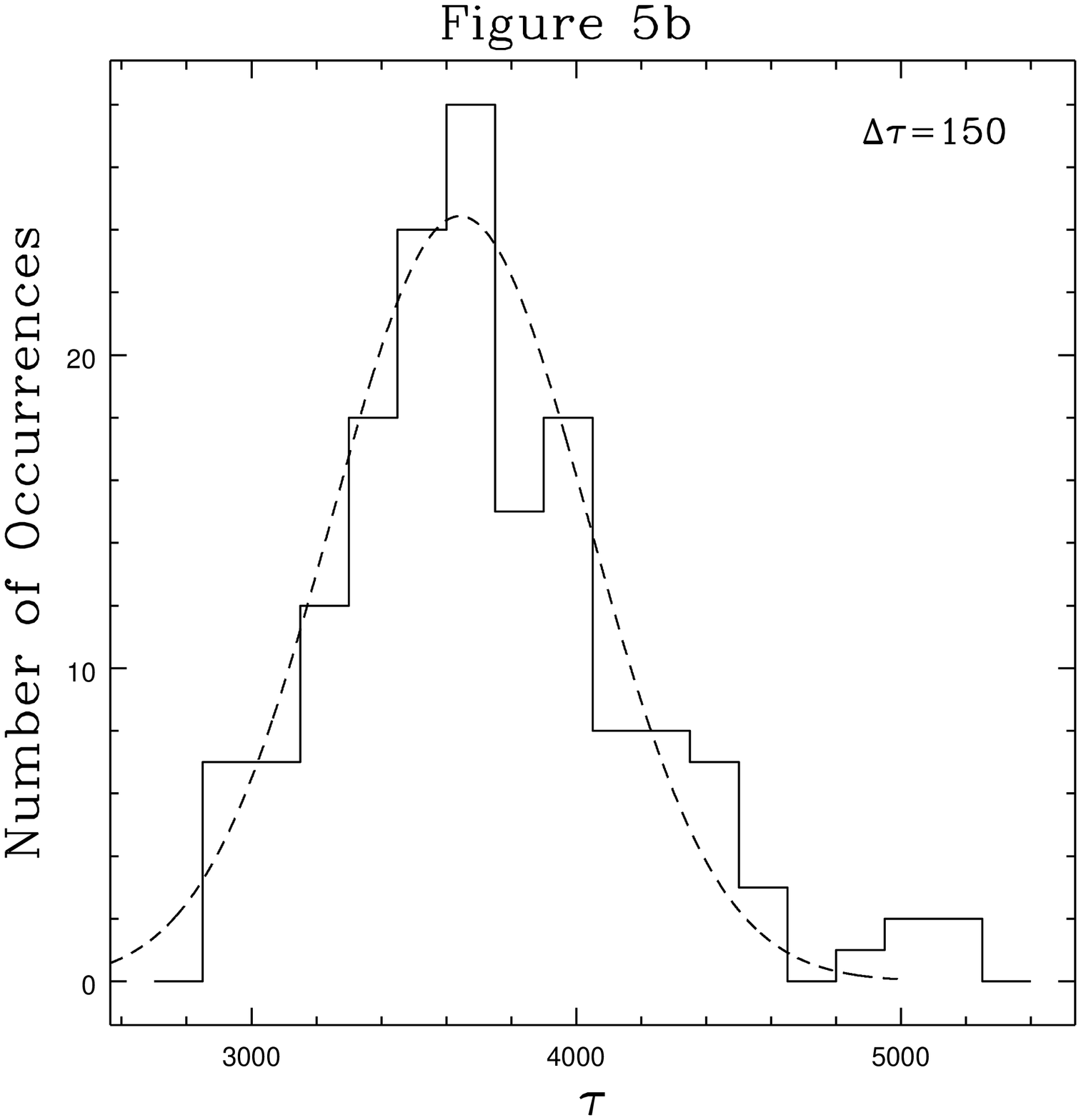,width=228pt}
\begin{figure}
	\caption{
		Histograms for
		(a) $R = 2.5$ and
		(b) $R = 3.0$, for
		$\gamma = 10^{-5}$ and $T=0.1$.
		$\Delta \tau$ indicates the bin size used in the
		specific cases.
	}
\end{figure}

\psfig{file=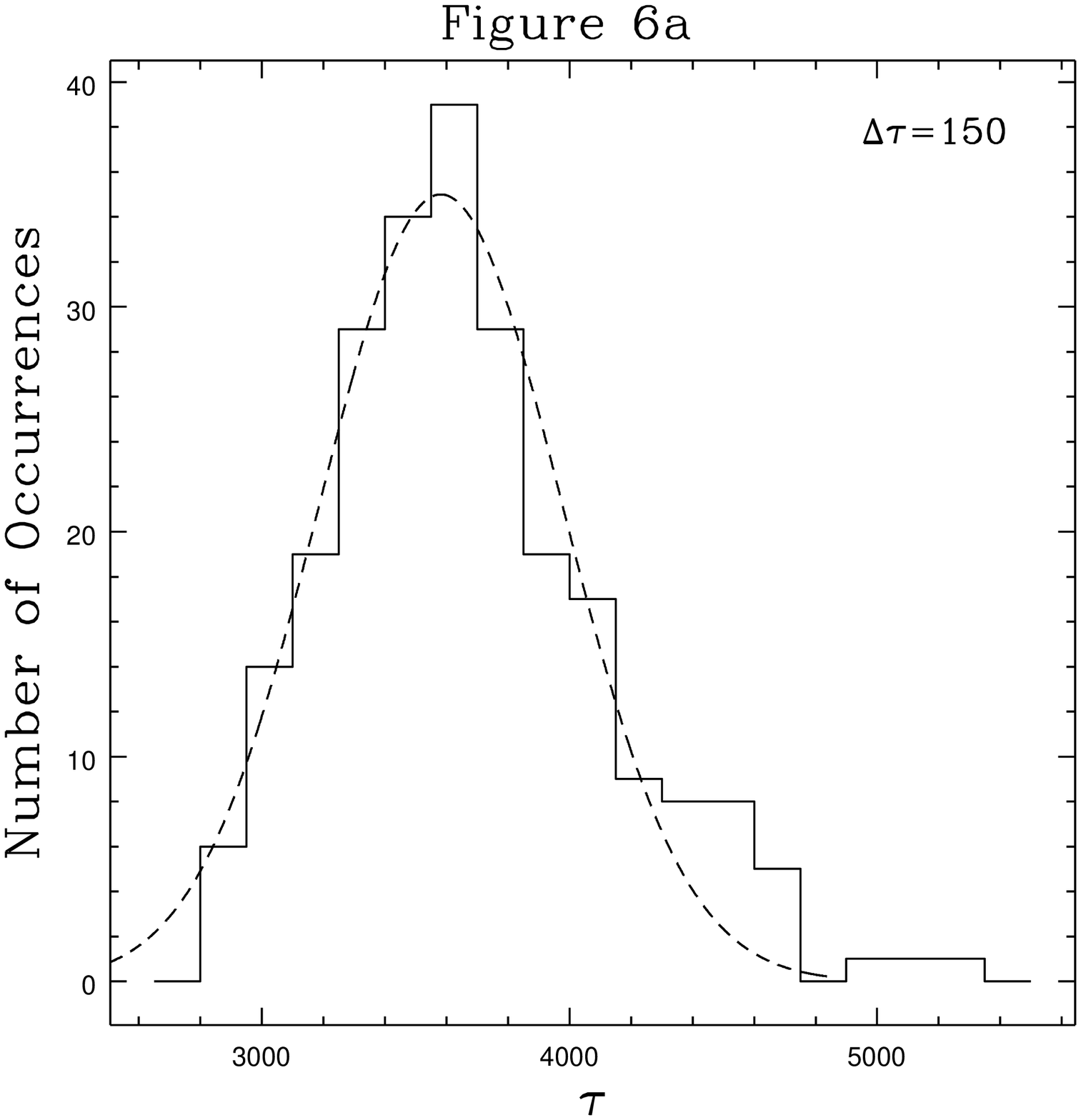,width=228pt}
\psfig{file=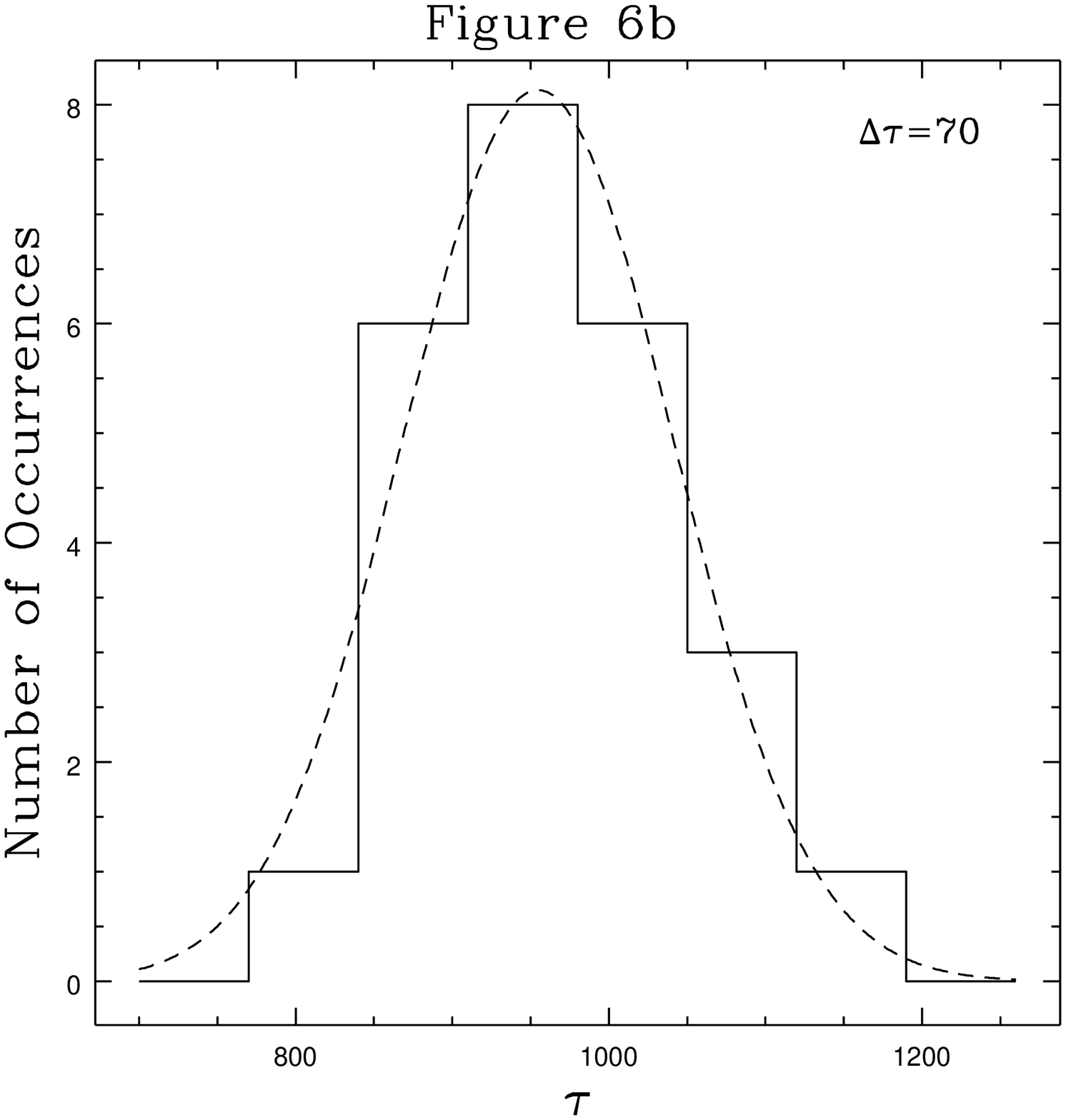,width=228pt}
\begin{figure}
	\caption{
		Histograms for
		(a) $\gamma = 10^{-5}$ and
		(b) $\gamma = 3 \times 10^{-4}$, for
		$R_{0} = 2.7$ and $T=0.1$.
		$\Delta \tau$ indicates the bin size used in the specific
		cases.
	}
\end{figure}

We can now extract the ensemble-averaged lifetime
by fitting a Gaussian to the various histograms, using a least-squares
approximation. Since the thermal noise producing the variation in the
lifetimes is generated from normal deviates, it is reasonable to assume
that the lifetimes will follow the central limit theorem, and produce
Gaussian distributed data as well, for a sufficiently large number of runs.
The mean gives the desired lifetime
and the variance gives the spread in the lifetime measurements. In order
to obtain the best possible fit, we used the Levenberg-Marquardt method
\cite{NUM-REC}.

The results for the lifetime as a function of radius and different values
of viscosity and temperature are shown in Figure 7.
It is clear that the 
coupling to the bath strongly suppresses the duration of the oscillon stage,
when it is at all present;
for values of viscosity
larger than $\gamma > 5 \times 10^{-4} m$, no oscillons could be found. Bubbles
quickly collapse, radiating their coherent energy into the incoherent
thermal bath. Since
$\gamma^{-1}$ is the typical thermalization time-scale, we obtain a
rough criterion for the existence of oscillons in the presence of a Markovian 
thermal bath,
\begin{equation}\label{existence}
\gamma^{-1} >  \tau_0~~,
\end{equation}
where $\tau_0$ is the lifetime of the cold oscillon of the same radius. Since 
typically $10^3\leq \tau_0 \leq 10^4$, this result is consistent with the
fact that the bath acts to destroy the coherence of the field 
configuration. 

\psfig{file=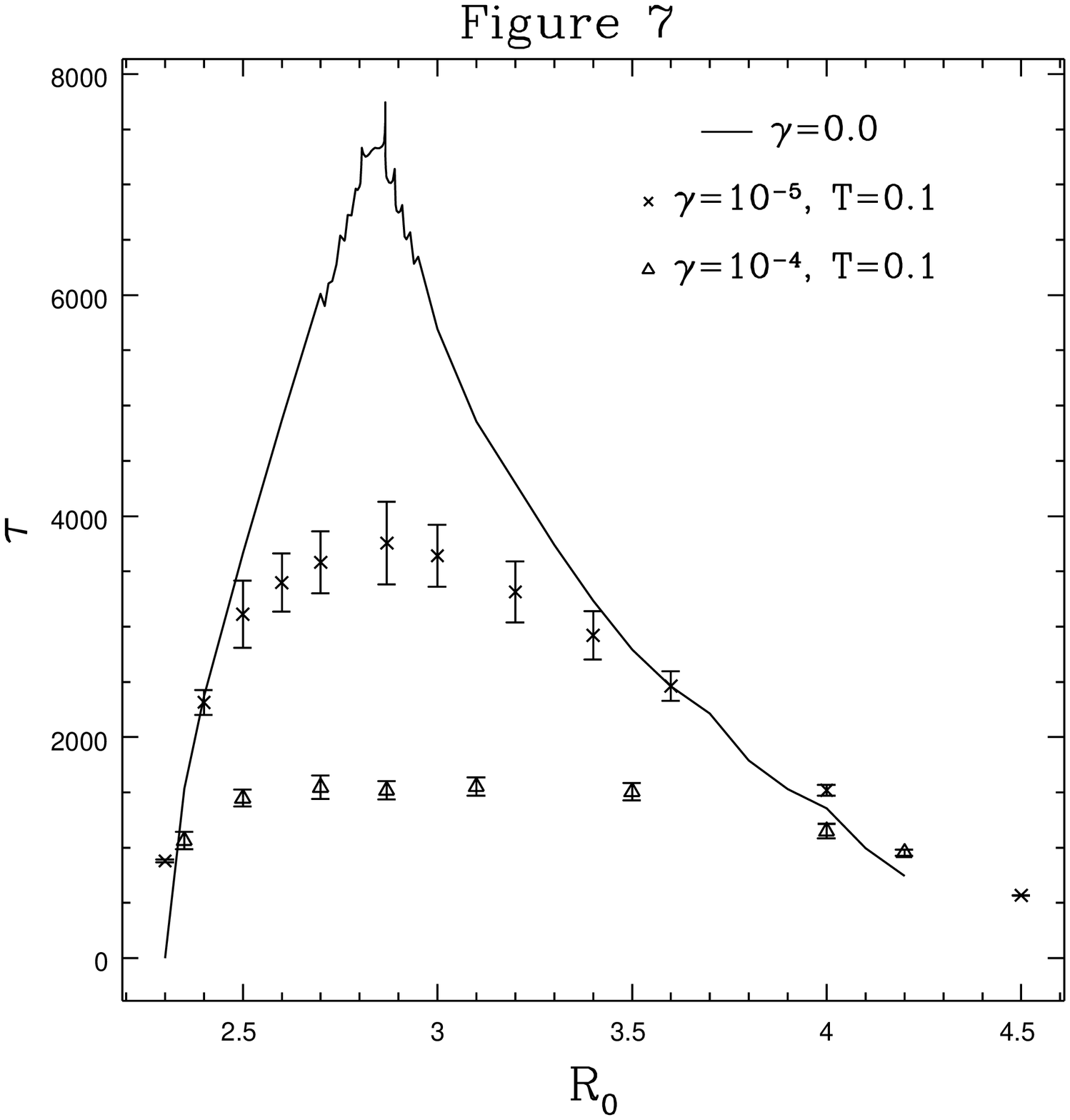,width=234pt}
\begin{figure}
	\caption{
		Oscillon lifetime vs. radius for constant viscosity and
		temperature.
	}		
\end{figure}

From Figure 7 we note that although the bath affects the longevity of
the oscillons, it does not appreciably affect
the range of radii for the initial bubble
settling into the oscillon stage; as in Refs. \cite{MG,CGM}, we still
obtain roughly $2.3 \lesssim R_0 \lesssim 4.5$ for the allowed range. Also,
the longest lived oscillons are the most affected, consistent with the above
condition, Eq. \ref{existence}.

In Figure 8 we display the results for lifetime vs. viscosity, for
$R_0=2.7m^{-1}$ and $T=0.1m/\lambda$. A least-squares fit produces the empirical
relation,
\begin{equation}
\tau(\gamma) = A\left [{{\gamma + B}\over {\gamma_0}}\right ]^{-C}~,
\end{equation}
where $A=2.0 \times 10^3,~B=4.0\times 10^{-6},~C=0.43$, and, $\gamma_0=
5.8\times 10^{-5}$. The error bars are given by the variance of the Gaussian 
fit the histograms. $\gamma_0$ sets the scale for the existence of oscillons.

\psfig{file=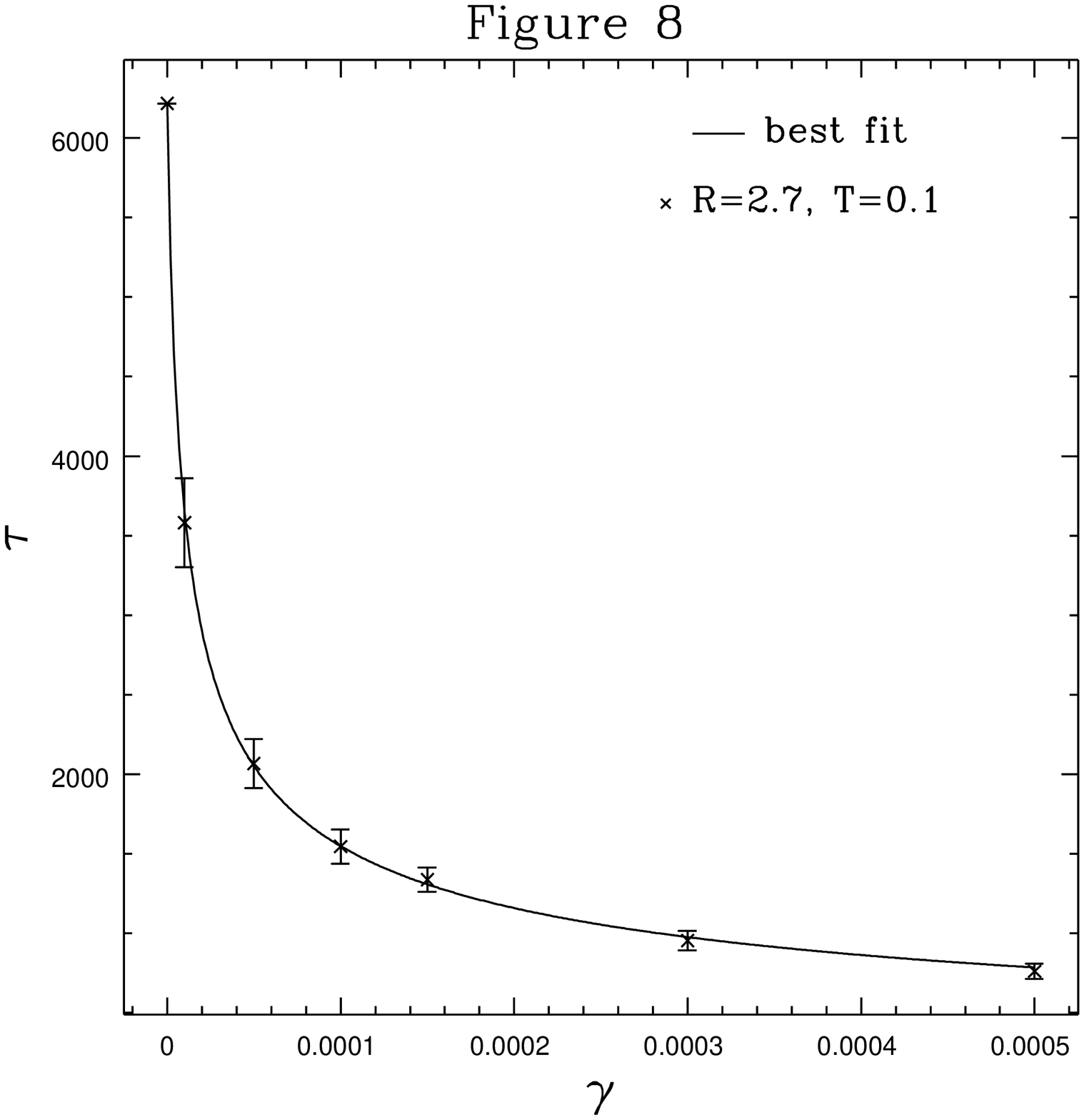,width=234pt}
\begin{figure}
	\caption{
		Oscillon lifetime vs. viscosity for
		constant radius and temperature.
		The equation providing best fit is
		$A \left( \frac{ \gamma + B }{\gamma_{0}} \right)^{-C}$
		where
		$A = 2.0 \times 10^{3}$,
		$B = 4.0 \times 10^{-6}$,
		$C = 0.43$, and
		$\gamma_{0} = 5.8 \times 10^{-5}$.
	}
\end{figure}

\psfig{file=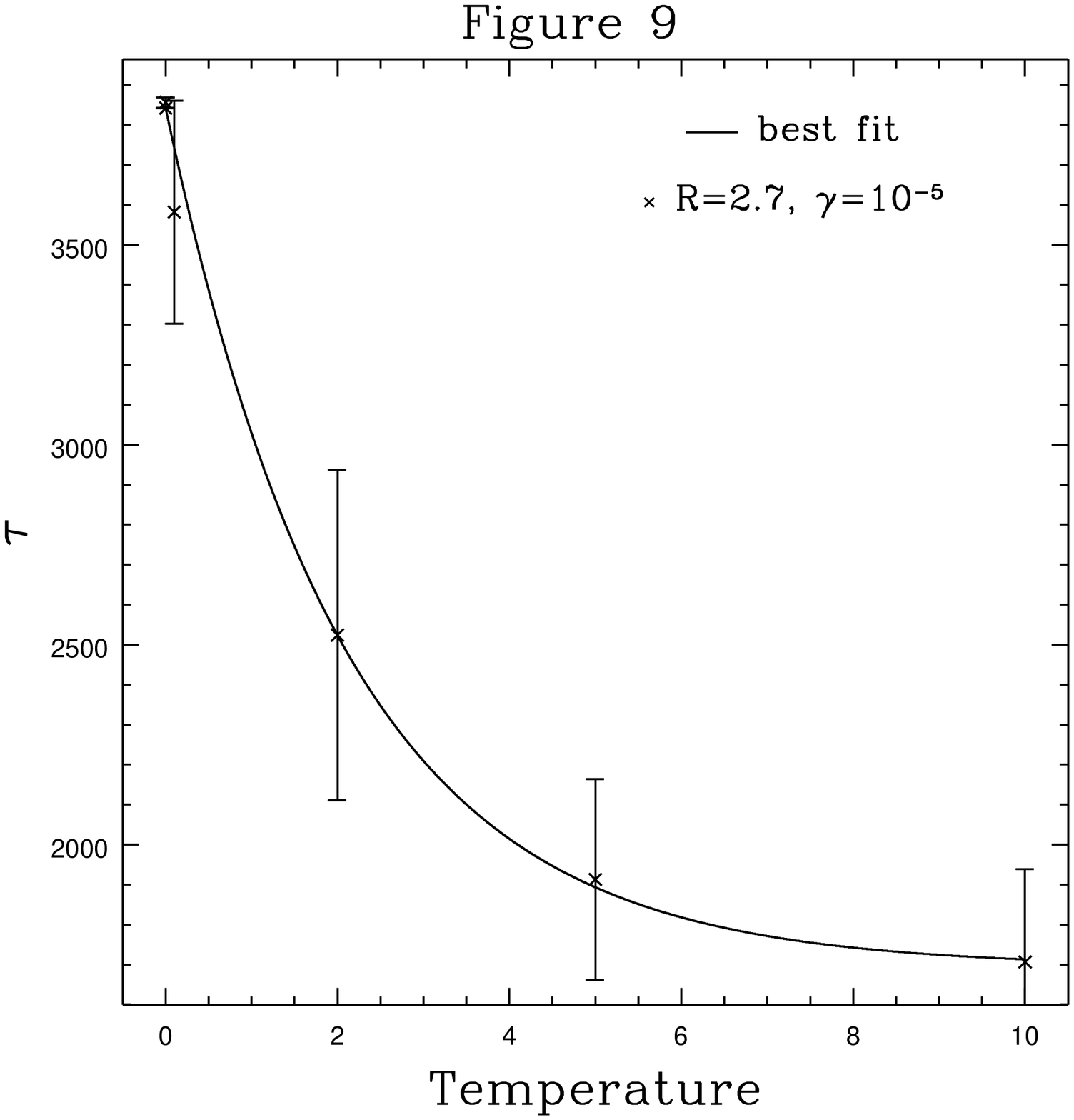,width=234pt}
\begin{figure}
	\caption{
		Oscillon lifetime vs. temperature for
		constant radius and viscosity.
		The equation providing best fit is
		$A \exp \left( - \frac{T}{T_{0}} \right) + B$
		where
		$A = 2.1 \times 10^{3}$,
		$B = 1.7 \times 10^{3}$, and
		$T_{0} = 2.1$.
	}
\end{figure}

Finally, in Figure 9 we display the results of lifetime vs. temperature,
for $R_0=2.7m^{-1}$ and $\gamma = 10^{-5}$. It is clear that the lifetime is
less sensitive to temperature then to viscosity. Again, an empirical fit
can be obtained,
\begin{equation}
\tau(T)= A{\rm exp}\left (-{T\over {T_0}}\right )+B~,
\end{equation}
where $A=2.1 \times 10^3,~B=1.7\times 10^3$, and $T_0=2.1$. For higher 
temperatures, it becomes very difficult to extract meaningful results, as the
signal to noise ratio sharply decreases.

\section{Conclusions}

We investigated the evolution of collapsing bubble-like configurations
in the presence of a thermal bath. The bubbles were assumed to have Gaussian
profile, and the dynamics were modelled by a generalized Langevin equation
with a Markovian thermal bath. In the absence of a thermal bath, it is possible
for these configurations to settle into what is known as an oscillon
stage, a nearly non-dissipative configuration which is thus 
extremely long-lived. 

We showed that the presence of a thermal bath affects the existence and 
longevity of the oscillon stage rather strongly. For values of the viscosity
coefficient, $\gamma > 5 \times 10^{-4}m$, 
no oscillon stage develops, independent
of temperature. This is naively analogous to an overdamped harmonic
oscillator. For smaller values of $\gamma$, an oscillon stage develops,
but its lifetime is suppressed. This result can be related to the fact
that the time-scale for thermalization is $\sim \gamma^{-1}$, while in the
absence of a heat bath the typical lifetime of the oscillon stage is $\tau_0\sim
10^3~ - ~ 10^4m^{-1}$. This is analogous to an underdamped harmonic oscillator,
where the decay time-scale is $\gamma^{-1}$. When the decay time-scale is
of order the lifetime of the oscillon, the bath suppresses the duration of
the oscillon stage. For even smaller values of $\gamma$, the oscillon 
approaches similar lifetimes
as in the cold oscillon case. Thus, the relevance of oscillons
in different contexts in which a thermal bath is present will depend on the
strength of the coupling between the field and the thermal bath, here modelled
by $\gamma$. As a rough rule, oscillons are present whenever 
$\gamma^{-1} > \tau_0$. 

Our discussion focused on Markovian thermal baths. However, given the
nature of interacting field theories, it is possible
that the bath will have more complicated properties such as spatio-temporal
correlations which are nonlocal, and/or couplings which are multiplicative
as oppose to additive. (For example, terms of the form $\xi\phi$ in the 
equation of motion \cite{GR}.) It remains to be seen if these nonlocal
effects will act as an additional source of coherence for fields. That being
the case, coupling to more general thermal baths 
may have very different consequences for the existence and longevity of
oscillons.

\acknowledgments

We would like to thank Hans-Reinhard M\"uller for many useful discussions
and assistance with the early development of the numerics.
MG was partially supported at Dartmouth by the National Science Foundation
through a  Presidential Faculty Fellows
Award no. PHY-9453431 and by a NASA grant no. NAGW-4270.

\end{multicols}
\end{document}